# Topology of nuclear reaction networks of interest for astrophysics

Carlos A. Bertulani[*]

*Department of Physics and Astronomy, Texas A&M University-Commerce, Commerce TX 75429, USA*

Our understanding of the observed elemental abundance in the universe, stemming from nuclear reactions during the big bang or from nucleosynthesis within stellar environments, requires theoretical analyses based on multidimensional nucleosynthesis calculations involving hundreds of nuclei connected via thousands of nuclear processes. Up to recently, full nucleosynthesis network calculations remained computationally expensive and prohibitive. A recent publication by a Chinese group led by YuGang Ma [1] has proved that advanced computational algorithms developed in the last decade for the purpose of studying complex networks are paving the way to finally accomplish this ultimate goal of nuclear astrophysics.

In the last two decades, parallel computing has been regarded as the main permitting factor to accomplish more precise and intensive simulations in nuclear astrophysics. However, even with code parallelization and with use of exaflop computers—equivalent to one quintillion floating- point computations ("flops") per second, where a flop equals two 15-digit numbers multiplied together—the challenges are daunting. Very often a huge amount of time is spent on communications between computer cores which, together with a limited number of isotopes in parts of the nuclear network, yield much worse performances of the parallelized mode compared to the one-core sequential version of the code. In order words, nucleosynthesis is a complex phenomenon, requiring a study of complex networks involving elements of topology, which is defined by how constituent parts of the network are interrelated or arranged.

Complex networks display non-trivial topological features beyond those appearing in systems formed by lattices or by random probability distributions. High clustering effects and power law distributions are common attributes pertaining to complex networks. In the last two decades the field of complex networks has become very active inspired by the need to understand timely problems in computer networks, biological networks, social network and other problems of human interest. Writing in *SCIENCE CHINA Physics, Mechanics & Astronomy*, Liu et al. [1] have found that nuclear reaction networks for nucleosynthesis calculations are amenable to many topological properties that have not yet been explored and that facilitate future theoretical and experimental studies in nucleosynthesis. The authors have accommodated existing information on 8048 nuclei and 82851 nuclear reactions in their nuclear reaction network computer codes. Such information, or data, have been compiled and are publicly available in the JINA nuclear database [2]. The foundation of their model is to separate the nuclear reactions into four kinds: (1) the n-layer, (2) p-layer, (3) h-layer and (4) r-layer, where each "layer" represents the kind of particle involved in the nuclear reaction. The four layers correspond to the reactions involving neutrons (n), protons (p) and helium (h) interacting with nuclei, whereas the remaining (r) layer includes all other nuclear reactions and decays.

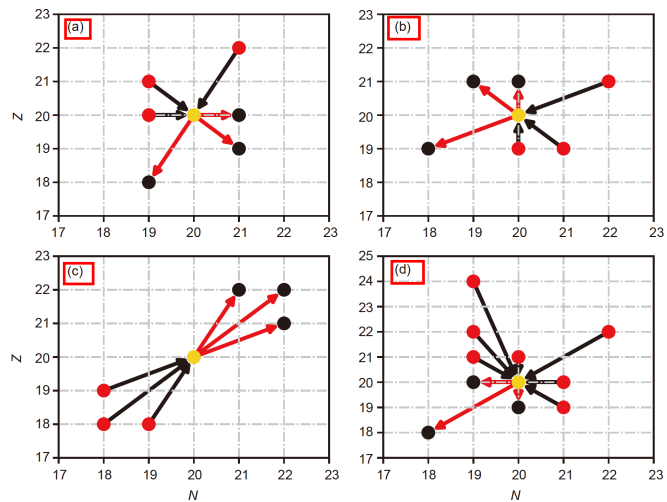

**Figure 1** (Color online) The topologies pertaining to the (a) n, (b) p, (c) h and (d) r layers of the $^{40}$Ca's nucleus at the nuclear landscape. The network is built on the basis of the substrate-product method.

Figure 1 shows an example of the nuclear reaction network topological structures with the four layers for the $^{40}$Ca nucleus denoted by golden point, in which the red arrow and the black arrow represent the consumption and the production of $^{40}$Ca, respectively. The complex network involving all nuclei and reactions is mapped onto topological structures using the substrate-product method which treats reactants and products as nodes and connects them if they are in the same reaction equation between the n, p, h and r multi-layers. The substrate-product method has been benchmarked and validated in previous studies [3].

The next step involves the analysis of the "motif" structure of the nuclear reaction network. By motif one means the identification of the most frequent reaction patterns of interconnections occurring in different nodes, with each node meaning one nucleus. These frequencies are correlated and the degree of correlation is assigned to each case. It was found out that nuclei in the n, p, and h layers have a similar degree value distribution, except for the nuclei containing more than about 82 protons. For such heavy nuclei, the distribution is different because of the neutron reactions, impacting more the n-layer nuclei than the p- and h-ones. The situation changes dramatically for the r-layer which displays a complex degree value due to the variety of reactions such as proton and neutron emission by photo-absorption, nuclear decay by beta-radiation, and so on. All networks can be represented as graphs—structures composed of a set of objects for which some object pairs are correlated—containing a large variety of subgraphs, or patterns. Network motifs are recurrent and built of statistically significant subgraphs. A motif provides a fine-grained description of the network.

The pioneer work of the Chinese group [1,4] on nuclear reaction networks of relevance for nucleosynthesis has unraveled that the n, p and h layers in nuclear reaction networks possess a regular structure, with a repetitive superposition of motifs. Another important finding is that the r-layer network is more complicated, having multiple motifs, but it is also a relief to notice that it belongs to the same category as the other layers. Another conclusion is that those motifs fade away for nuclei with the proton or neutron number around and above 82. Future work by the group will focus on improving mapping methods, incorporating the dynamics on the network and a more in-depth analysis of each layer. Topological descriptions of nuclear reaction networks for the purpose of calculating elemental abundance in the universe will shorten the computation time and allow for a more efficient use of supercomputers to achieve the accuracy needed to reproduce astronomical observations.

**References**

1. H. L. Liu, D. D. Han, Y. G. Ma, and L. Zhu, Sci. China-Phys. Mech. Astron. **63**, 112062 (2020).
2. R. H. Cyburt, A. M. Amthor, R. Ferguson, Z. Meisel, K. Smith, S. Warren, A. Heger, R. D. Hoffman, T. Rauscher, A. Sakharuk, H. Schatz, F. K. Thielemann, and M. Wiescher, Astrophys. J. Suppl. Ser. **189**, 240 (2010).
3. P. Holme, J. R. Soc. Interface **6**, 1027 (2009).
4. L. Zhu, Y. G. Ma, Q. Chen, and D. D. Han, Sci. Rep. **6**, 31882 (2016), arXiv: 1608.07812.